\documentclass{aipproc}
\layoutstyle{6x9}
\begin{document}

\title{Diffusive phase transitions in ferroelectrics and
antiferroelectrics}

\author{S. A. Prosandeev}
{
  address={Physics Department, Rostov State University, 5 Zorge St.,
344090 Rostov on Don, Russia}  }
\author{I. P. Raevski}
{
 address={Physics Department, Rostov State University, 5 Zorge St.,
344090 Rostov on Don, Russia}  }
\author{U. V. Waghmare}{
  address={J Nehru Centre for Advanced Scientific Research, Bangalore,
India} }

\begin{abstract}
In this paper, we present a microscopic model for heterogeneous
ferroelectric and an order parameter for relaxor phase. We write a
Landau theory based on this model and its application to
ferroelectric PbFe$_{1/2}$Ta$_{1/2}$O$_3$~ (PFT) and
antiferroelectric NaNbO$_3$:Gd. We later discuss the coupling
between soft mode and domain walls, soft mode and quasi-local
vibration and resulting susceptibility function.
\end{abstract}
\maketitle

\begin{figure}
  \includegraphics[height=.4\textheight]{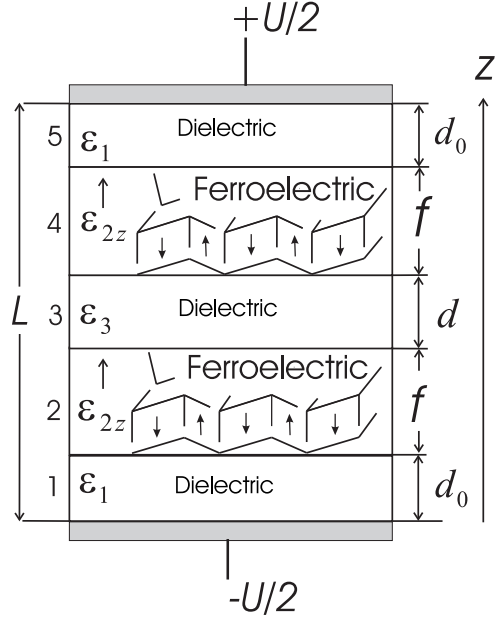}
  \caption{A model of inhomogeneous ferroelectrics [2].}
\label{model}
\end{figure}

A simple model ferroelectric thin film with dead layers was
recently analyzed by Bratkovsky and Levanyuk \cite{Bratkovsky}
resulting in an analytic solution. Based on this study, we
proposed a microscopic model for an inhomogeneous ferroelectric
consisting of ferroelectric slabs sandwiched between dielectric
interfacial layers \cite{model}. Low energy solutions of these
models reveal that the ferroelectric slabs break into alternating
domains (Fig. \ref{model}) with zero total macroscopic
polarization. The size of the domains was shown to depend only on
the relative total width of the dielectric and ferroelectric
regions in the direction of the field \cite{model}. The nanodomain
structure appears cooperatively and its origin lies in the
reduction of depolarization field \cite{Bratkovsky,model}.

The alternating polarization domains are accompanied by shear
strain owing to the electrostrictive coupling. For the case of
domains alternating along (110) direction, we obtained atomic
displacements corresponding to polarization and strain fields and
calculated diffused scattering intensity \cite{model}, which
agrees well with recent neutron scattering data \cite{Hirota}. We
suggest that the relaxor phase corresponds to a finite value of
the order parameter describing these alternating domains.

The fluctuations of the order parameter introduced above are
conjugated with the field which we will call $H_0$. We will assume
the existence of quenched field $H_0$~ below Burns temperature. We
use the notation $\eta$~ for the relaxor order parameter (the
magnitude of polarization inside a nanodomain) and write the
Landau expansion:

\begin{eqnarray}
F=F_0 + \frac{1}{2} \alpha P^2 + \frac{1}{4} \beta P^4 +
\frac{1}{6} \gamma P^6 +  \nonumber \\ \frac{1}{2} A \eta^2 +
\frac{1}{4} B \eta^4 +\frac{1}{6} C \eta^6 -H_0 \eta + \frac{1}{2}
\lambda P^2 \eta^2
\end{eqnarray}
where $\alpha = \alpha_0 (T-T_{CW})$~ and $A=a(T-T_\eta)$.

At zero macroscopic polarization, the equilibrium condition with
respect to $\eta$~ is

\begin{equation}
A\eta + B \eta^3 +C \eta^5 - H_0 = 0 \label{vector}
\end{equation}
At high temperatures $\eta$~ vanishes as $(T-T_{\eta})^{-1}$. In
this limit, dielectric permittivity

\begin{equation}
\chi=\frac{1}{v(T-T_{CW})+\lambda \eta^2,} \label{chi}
\end{equation}
obeys a Curie-Weiss law. At $T=T_{\eta}$~ there is a deviation
from this law, and the dielectric permittivity peak is diffused
due to the coupling between the new order parameter appearing at
the phase transition and frozen conjugated fields.

We used the expression derived in order to describe the
diffuseness of the phase transition in relaxor PFT~ from the
para-phase to the relaxor phase with ferroelectric nano-regions.
We also took into account a low temperature (glass-type) phase
transition which results in a strong decrease of dielectric
permittivity below approximately 220 K. We fitted expressions
obtained to experimental data (Fig. 2) and got the best fit shown
by the solid line. We find that $T_{\eta}$~ is rather close to the
extrapolated high-temperature Curie-Weiss temperature. This
justifies that the ferroelectric relaxor order parameter, $\eta$,
is connected with local polarization, and can be regarded to a
wave of polarization described above.

We use the same expression (3) to treat the diffused phase transition
in an antiferroelectric NaNbO$_3$ doped with Gd, where the significance of
$\eta$~ now is an antiferrolectric order parameter.
The resulting fit to experimental data is rather good shown in Fig. 3.
The critical temperatures are indicated in the figure.
We also show in Fig. 3 the difference between the inverse dielectric permittivity
and high temperature Curie-Weiss law. It is seen that this
difference, at low temperatures, behaves linearly with temperature
while at the dielectric permittivity peak position it is diffused
in accord with the theory.

Experimental data \cite{Gehring} show that the ``waterfall
phenomenon'' exists in the whole temperature interval between the
freezing temperature and Burns temperature, implying that the soft
mode is strongly damped in the relaxor phase (note that this soft
mode is not the uniform ferroelectric). We believe this mechanism
to be connected with the interaction of the soft mode vibrations
with domain walls and with local dipoles (a mathematical
expression is similar to that obtained for the soft mode coupled
with microscopic dipoles \cite{Kleemann,Trepakov}).

\begin{figure}
  \includegraphics[height=.4\textheight]{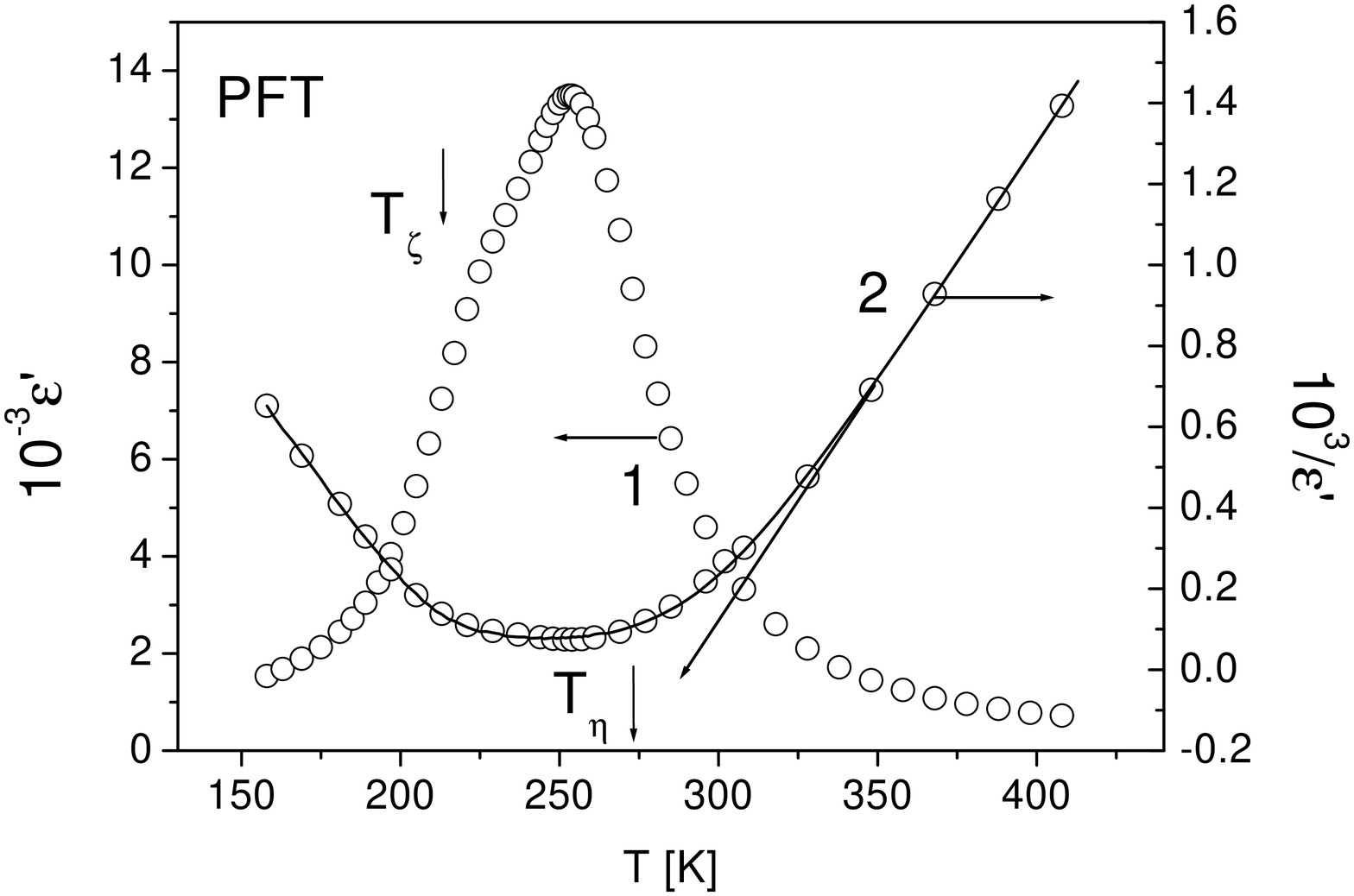}
  \caption{Temperature dependencies of $\varepsilon'$~ (1) and
$10^3/ \varepsilon' $~ (2) measured at 10$^6$~ Hz for PFT crystal.
Solid line is the best fit of the theory to experimental data.}
\label{PFT}
\end{figure}

The acoustic mode can have a dip due to its coupling with optical
mode \cite{Yamada} (Fig. 4a). We suggest that, due to disorder,
the deviation of the optical and acoustic modes from average is
different in different regions of the disordered crystal and one
has to introduce a distribution function of these deviations (Fig.
4b). We also consider the case when the optical mode has a dip
(Fig. 4c) if the interaction with the acoustic mode is not taken
into account (the first row in Fig. 4). In this case the optical
mode dispersion curve is: $\varepsilon=\omega_0^2 - kq^2 + bq^4 +
...$. We derived in this case that the correlation function
deviates from the Ornstein-Cernike expression by an oscillating
factor:

\begin{equation}
\chi (r) \sim \frac{1}{r} \sin(k_0 r) \exp(-k_c r) \label{chir}
\end{equation}
where $k_c$~ is inverse correlation length and $k_0$~ is the wave
vector of the susceptibility oscillations. In the $k$-space
susceptibility looks as

\begin{equation}
\chi(q) \sim \frac{a^2}{\left(q^2-k_m^2 \right)^2 + a^4}
\end{equation}
Here $k_m^2=k_0^2-k_c^2$~ and $a^2=2k_0 k_c$. These constants are
expressed in terms of parameters in the optical mode dispersion:
$\omega_0^2/b=k_m^4+a^4$~ and $k/b=2k_m^2$. The inverse
correlation length $k_c$ can be obtained from:

\begin{figure}
  \includegraphics[height=.4\textheight]{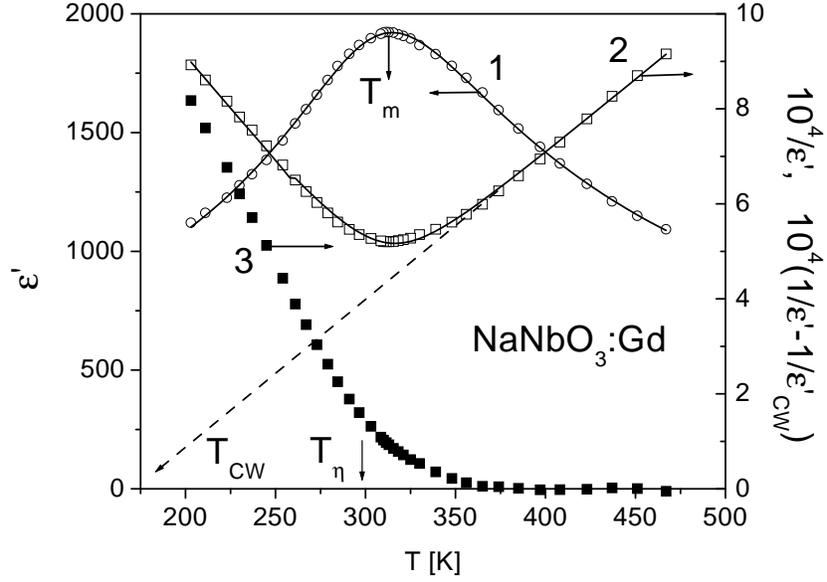}
  \caption{Temperature dependencies of $\varepsilon'$ (1) and $10^4/
\varepsilon' $ (2) measured at 10$^5$ Hz for
0.88NaNbO$_{3}$-0.12Gd$_{1 / 3 }$NbO$_{3}$~ crystal, and the
difference between the experimental $1/\varepsilon'$~ dependence
and the Curie-Weiss fit of $1 / \varepsilon'$~ (3). Solid lines
are the best fits of the theory to experimental data.}
\label{NaNbGd}
\end{figure}

\begin{equation}
k_c^2=\frac{1}{2}\left[\sqrt{
\frac{\omega_0^2}{b}}-\frac{k}{2b}\right] \label{kc}
\end{equation}
where, at $\omega_0^2 > k >0$, $k_c$~ is real. Otherwise, it is
imaginary, and there appear harmonic beatings.

A general form of the Hamiltonian taking into account the
interaction between the acoustic ($u$) and optical ($x$)
displacements (and a dip in the optical mode) is:

\begin{equation}
H_{harm}(q) = \frac{1}{2}u_{-q}A(q)u_q + u_{-q}V x_q + x_{-q}V^*
u_q + \frac{1}{2} x_{-q} B(k) x_q  \label{Hamiltonian}
\end{equation}
where, at small $q$,

\begin{eqnarray}
A(q) &=& a q^2 \\ B(k) &=& \omega_0^2 - bq^2 +cq^4 \\ V(q) &=& v_1
q^2 + iv_2Pq
\end{eqnarray}
Here P is polarization; The imaginary part of the coupling term
appears in symmetry broken regions and in the regions where there
exists gradient of polarization, that is at the boundaries of the
domains or/and polar regions. This coupling leads to a
``repulsion'' of the acoustic and optical modes as it is shown in
Fig. 4.

We also will discuss the correlation of polarization in the case
of a finite volume of the polar region.  Let us write dielectric
permittivity of a finite uniform volume $V$:

\begin{figure}
  \includegraphics[height=.4\textheight]{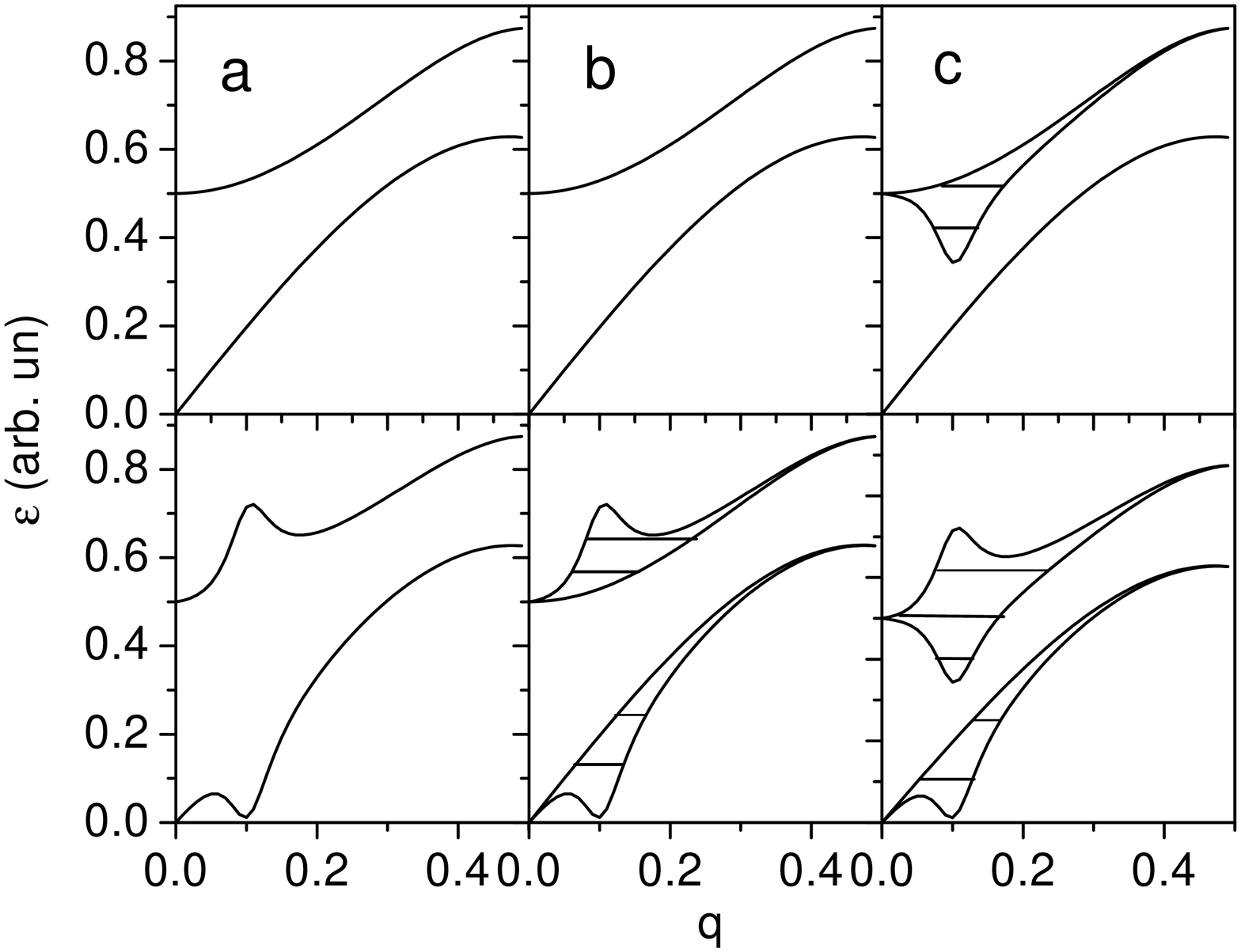}
  \caption{Model dispersion curves: a) the condensation of the modes
is due to mode-mode coupling, b) the same as (a) but with taking
into account disorder, c) the condensation of the modes is due to
both the polarization instability and mode-mode coupling; The
first row is without the mode-mode coupling, the second row is
with the mode-mode coupling; The dashed region arises because of
disorder.} \label{dispersion}
\end{figure}

\begin{equation}
\chi \sim \frac{T}{V^2}\int dVdV'\left<P(r)P(r')\right> \sim
\frac{T_{CW}}{\kappa V} min(r^2,r_c^2)
\end{equation}
where an Ornstein-Cernike correlator was used, $\kappa$~ being
the constant in front of $\left(\nabla P \right)^2 $~ in the
Landau fluctuation energy, and $r_c \sim \left(T-T_{CW}
\right)^{-1/2}$, in the first approximation. It is seen from this
expression that, at high temperatures, susceptibility behaves
according to the Curie-Weiss law; when the ferroelectric
correlation radius reaches the volume $V$~ size then
susceptibility saturates. The decrease in polarization at the
boundary of a polar region would smoothen this crossover.

We finally consider the Hamiltonian consisting of the part describing
the long-range ordered polarization $P_{\bf{q}}$~ (with a wave
vector \textbf{q}) and coupled with it local polar vibrations with
local polarization $P_l$:

\begin{eqnarray}
H&=&\frac{1}{2} \alpha_{\bf{q}} P_{\bf{q}}^2 + \frac{1}{2}
\alpha_l P_l^2 - \alpha_{{\bf{q}}l} P_{\bf{q}}P_l
+\frac{1}{4}\beta P_{\bf{q}}^4 +\frac{1}{6}\gamma P_{\bf{q}}^6
\nonumber \\ && +\frac{1}{4} B P_l^4 + \frac{1}{2} g P_{\bf{q}}^2
P_l^2 -E P_{\bf{q}} - E P_l
\end{eqnarray}

The susceptibility connected with quasilocal vibrations at zero
polarization $P_{\bf{q}}$~ can be found by an ordinary procedure
\cite{Kleemann}:

\begin{equation}
\chi_l = \frac{\chi_{l0}}{1-\alpha_{{\bf{q}}l}^2 \chi_{{\bf{q}}0}}
\end{equation}
where

\begin{eqnarray}
\chi_{l0}=\frac{1}{\alpha_l + 3 \beta P_l^2} \\ \chi_{{\bf{q}}0}=
\frac{1}{\alpha_{\bf{q}} + g P_l^2}
\end{eqnarray}

It is seen that the quasilocal vibrations become unstable when
$\alpha_{{\bf{q}}l}^2 \chi_{{\bf{q}}0} <1$. It implies that the
local vibrations satisfying this condition will freeze in and
will be arranged in space in accordance with the wave vector
\textbf{q}. These results are consistent with resent experimental
findings \cite{Dulkin} that there is a local transformation at
Burns temperature.

Partially supported by RFBR grants 01-03-33119 and 01-02-16029.


\begin{thebibliography}{999}

\bibitem{Bratkovsky} Bratkovsky A. M. and Levanyuk A. P.,  {\it Phys. Rev. Lett.}
\textbf{84} 3177; \textit{ibid} \textbf{86} 3642 (2001).

\bibitem{model}Raevski I. P., Prosandeev S. A., Waghmare U.,
Eremkin V. V., Smotrakov V. G., Shuvaeva V. A., cond-mat/0208116.
\bibitem{Hirota}Hirota K., Ye Z. -G., Wakimoto S., Gehring P. M., and Shirane
G.,
{\it Phys. Rev. B} \textbf{65} 104105 (2002).
\bibitem{Gehring}Gehring P. M., Wakimoto S., Ye Z. -G. and Shirane G.,  {\it
Phys. Rev. Lett.} \textbf{87} 277601-1 (2001).
\bibitem{Kleemann}Prosandeev S. A., Kleemann W. and Dec J., {\it J Phys:
Condens Matter} \textbf{13} 5957 (2001).
\bibitem{Trepakov}Prosandeev S. A., Trepakov V. A., Savinov M. E.,
Jastrabik L. and Kapphan S. E., {\it J. Phys: Condens. Matter}
\textbf{13} 9749 (2001).
\bibitem{Yamada}Yamada Y., Takakura T., cond-mat/0209573.
\bibitem{Dulkin}
Dul'kin E., Raevski I. P., and Emelyanov S. M., {\it Phys Sol St}
{\bf 45} 158 (2003).
\end{thebibliography}
\end{document}